# Mathematical structure of three - dimensional (3D) Ising model

Zhi-dong Zhang

Shenyang National Laboratory for Materials Science, Institute of Metal Research, Chinese Academy of Sciences, 72 Wenhua Road, Shenyang, 110016, P.R. China

e-mail: zdzhang@imr.ac.cn

An overview of the mathematical structure of the three-dimensional (3D) Ising model is given, from the viewpoints of topologic, algebraic and geometric aspects. By analyzing the relations among transfer matrices of the 3D Ising model, Reidemeister moves in the knot theory, Yang-Baxter and tetrahedron equations, the following facts are illustrated for the 3D Ising model: 1) The complexified quaternion basis constructed for the 3D Ising model represents naturally the rotation in a $(3 + 1)$ - dimensional space-time, as a relativistic quantum statistical mechanics model, which is consistent with the 4-fold integrand of the partition function by taking the time average. 2) A unitary transformation with a matrix being a spin representation in $2^{n \cdot l \cdot o}$-space corresponds to a rotation in $2n \cdot l \cdot o$-space, which serves to smooth all the crossings in the transfer matrices and contributes as the non-trivial topologic part of the partition function of the 3D Ising model. 3) A tetrahedron relation would ensure the commutativity of the transfer matrices and the integrability of the 3D Ising model, and its existence is guaranteed also by the Jordan algebra and the Jordan-von Neumann-Wigner procedures. 4) The unitary transformation for smoothing the crossings in the transfer matrices changes the wave functions by complex phases $\phi_x$, $\phi_y$, and $\phi_z$. The relation with quantum field and gauge theories, physical significance of weight factors are discussed in details. The conjectured exact solution is compared with numerical results, and singularities at/near infinite temperature are inspected. The analyticity in $\beta = 1/(k_B T)$ of both the hard-core and Ising models has been proved for $\beta > 0$, not for $\beta = 0$. Thus the high-temperature series cannot serve as a standard for judging a putative exact solution of the 3D Ising model.

## I. INTRODUCTION

The Ising model, which serves as a model system for understanding critical phenomena with profound physical significances, has been intensively investigated. The two-dimensional (2D) Ising model was solved explicitly by Onsager.[1] In ref. [2], we reported a putative exact solution of the 3D simple orthorhombic Ising model based on two conjectures repeated here:

*Conjecture 1:* The topologic problem of a 3D Ising system can be solved by introducing an additional rotation in a four-dimensional (4D) space, since the knots in a 3D space can be opened by a rotation in a 4D space. One can find a spin representation in $2^{n \cdot l \cdot o}$-space for this additional rotation in $2n \cdot l \cdot o$-space. Meanwhile, the transfer matrices $V_1$, $V_2$ and $V_3$ also have to be represented and rearranged in the $2n \cdot l \cdot o$-space.

*Conjecture 2:* The weight factors $w_x$, $w_y$ and $w_z$ on the eigenvectors represent the contribution of $e^{i\frac{2t_x\pi}{n}}$, $e^{i\frac{2t_y\pi}{l}}$ and $e^{i\frac{2t_z\pi}{o}}$ in the 4D space to the energy spectrum of the system.

After publication of ref. [2], two rounds of exchanges of Comments / Responses / Rejoinders appeared in 2008-2009.[3-8] The main objections of these Comments and Rejoinders[3,5,6,8] are summarized briefly as follows: The conjectured solution disagrees with low-temperature and high-temperature series, while the convergence of the high-temperature series has been rigorously proved. There are some problems with weight factors $w_y$ and $w_z$ (such as they are first defined to be real values, but actually can be complex, and the different weights are taken for infinite and finite temperature). There is a technical error in eq. (15) of ref. [2] for the application of the Jordan–Wigner transformation. The main rebuttals in my Responses[4,7] are also summarized briefly here: The objections in the Comments/Rejoinders[3,5,6,8] are limited to the outcome of the calculations and there were no comments on the topology-based approach underlying the derivation. All the well-known theorems for the convergence of the high-temperature series are proved only for $\beta$ (= $1/k_B T$) > 0, not for infinite temperature ($\beta$ = 0). Exactly infinite temperature has been never touched in these theorems, since there exists a singularity at $\beta$ = 0. From another angle of view, the conjectured solution and its low- and high-temperature expansions are supported by the mathematical theorems for the analytical behavior of the 3D Ising model. Furthermore, the failure in reproducing term by term the low-temperature expansion which is divergent does not disqualify the new approach for dealing with knots by means of an extension into a fourth dimension. The weight factors can be generalized to be complex. The error of Equation (15) in ref. [2] does not affect the validity of the putative exact solution, since the conjectures serve for solving the topological problems existing in that equation.

Further progresses have been made in ref. [4,7,9-12]. In ref. [4], Conjecture 1, regarding the additional rotation, was understood as performing a unitary transformation for smoothing all the crossings of the knots, while the weight factors in Conjecture 2 were interpreted as novel topologic phases. More recently, the algebraic part of the quaternion approach used in ref. [2] was reformulated in terms of the quaternionic sequence of Jordan algebras to look at the geometrical aspects of simple orthorhombic Ising lattices,[9] and fractals and chaos related to these 3D Ising lattices were investigated.[10-12]

In this work, we represent an overview of the mathematical structure of the 3D Ising model from the viewpoints of topologic, algebraic and geometric aspects,* and attempt to bridge the gaps in the conjectures. This paper is arranged as follows: In Section II, the Hamiltonian, transfer matrixes, boundary conditions, and the consequences of the two conjectures are introduced briefly (and the technical errors in Eqs. (15) and (16) of ref. [2] have been corrected). In Section III, an overview of the mathematical structure of the 3D Ising models is given, where algebraic, topologic and geometric aspects are all related together. Section IV discusses the relativistic quantum statistical mechanics and its relation with quantum field and gauge theories, physical significance of weight factors, and singularities at/near infinite temperature. A comparison of the exact solution with numerical results is also present. Conclusions are given in Section V, together with some opening problems.

---

*This paper is partially based an invited talk, "Outlook on the mathematical structure of the three-dimensional (3D) Ising model" at Hypercomplex Seminar 2012: (Hyper)Complex Function Theory, Regression, (Crystal) Lattices, Fractals, Chaos, and Physics, at Będlewo, Poland, July 08-15, 2012.

## II. HAMILTONIAN, TRANSFER MATRIXES AND CONSEQUENCES OF THE CONJECTURES

The Hamiltonian of the 3D Ising model on simple orthorhombic lattices is written as:[2]

$$\widehat{H} = -J\sum_{\tau=1}^{n}\sum_{\rho=1}^{m}\sum_{\delta=1}^{l} s_{\rho,\delta}^{(\tau)} s_{\rho,\delta}^{(\tau+1)} - J'\sum_{\tau=1}^{n}\sum_{\rho=1}^{m}\sum_{\delta=1}^{l} s_{\rho,\delta}^{(\tau)} s_{\rho+1,\delta}^{(\tau)} - J''\sum_{\tau=1}^{n}\sum_{\rho=1}^{m}\sum_{\delta=1}^{l} s_{\rho,\delta}^{(\tau)} s_{\rho,\delta+1}^{(\tau)}. \qquad (1)$$

The partition function becomes:[2]

$$Z = trace(T(V))^m = trace(V_3 V_2 V_1)^m \qquad (2)$$

with the transfer matrices $V_1$, $V_2$ and $V_3$ being:

$$V_3 = \exp\{K''\cdot\sum_{r=1}^{n}\sum_{s=1}^{l} s'_{r,s}\, s'_{r,s+1}\} \equiv \exp\{K''\cdot A''\}, \qquad (3)$$

$$V_2 = \exp\{K'\cdot\sum_{s=1}^{l}\sum_{r=1}^{n} s'_{r,s}\, s'_{r+1,s}\} \equiv \exp\{K'\cdot A'\}, \qquad (4)$$

$$V_1 = (2\sinh 2K)^{\frac{n\cdot l}{2}} \cdot \exp\{K*\cdot\sum_{s=1}^{l}\sum_{r=1}^{n} C_{r,s}\}. \qquad (5)$$

Here we introduced variables $K \equiv J/k_BT$, $K' \equiv J'/k_BT$ and $K'' \equiv J''/k_BT$ instead of J, J' and J''.[2] K* is defined by $e^{-2K} \equiv \tanh K*$.[1,2,13] We introduced $2^{n\cdot l}$-dimensional quaternion matrices $s''_{r,s}$, $s'_{r,s}$ and $C_{r,s}$:

$$s''_{r,s} = C_{1,s}C_{2,s}\cdots C_{r-1,s}Q_{r,s} \equiv 1\otimes 1\otimes\ldots\otimes 1\otimes s''\otimes 1\otimes\ldots\otimes 1, \qquad (6a)$$

$$s'_{r,s} = C_{1,s}C_{2,s}\cdots C_{r-1,s}P_{r,s} \equiv 1\otimes 1\otimes ...\otimes 1\otimes s'\otimes 1\otimes ...\otimes 1, \quad (6b)$$

$$C_{r,s} = iP_{r,s}Q_{r,s} \equiv 1\otimes 1\otimes ...\otimes 1\otimes C\otimes 1\otimes ...\otimes 1, \quad (6c)$$

There are n·l factors in each direct-product, with **s''**, **s'** and **C** appearing in the (r,s)-th position. **s''**, **s'** and **C** are generators of the Pauli spin matrices:

$$s'' \equiv \begin{bmatrix} 0 & -1 \\ 1 & 0 \end{bmatrix}, s' \equiv \begin{bmatrix} 1 & 0 \\ 0 & -1 \end{bmatrix}, C \equiv \begin{bmatrix} 0 & 1 \\ 1 & 0 \end{bmatrix}, 1 = \begin{pmatrix} 1 & 0 \\ 0 & 1 \end{pmatrix}. \quad (7)$$

$P_{r,s}$ and $Q_{r,s}$ are related with a set of 2n·l-quantities $\Gamma_k$:

$$\Gamma_{2r-1} \equiv C\otimes C\otimes \cdots \otimes s\otimes 1\otimes 1\otimes \cdots \equiv P_r,$$

$$\Gamma_{2r} \equiv -C\otimes C\otimes \cdots \otimes isC\otimes 1\otimes 1\otimes \cdots \equiv Q_r, \quad 1\leq r\leq nl, \quad (8)$$

where n·l factors appear in each product; **s** or **isC** appears in the rth place. The $\Gamma_k$ are $2^{n\cdot l}$-dimensional matrices, which obey the commutation rules

$$\Gamma_k^2 = 1, \Gamma_k\Gamma_l = -\Gamma_l\Gamma_k, (1\leq k, l, \leq 2n\cdot l). \quad (9)$$

Following the works of Kaufman[13] and Lou and Wu[14] and also discussion in refs. [6,7], we can give the partition function

$$Z = (2\sinh 2K)^{\frac{m\cdot n\cdot l}{2}}\cdot trace(V_3V_2V_1)^m \equiv (2\sinh 2K)^{\frac{m\cdot n\cdot l}{2}}\cdot \sum_{i=1}^{2^{n\cdot l}}\lambda_i^m \quad (10)$$

while matrices $V_1$, $V_2$ and $V_3$ can be represented by performing the Jordan-Wigner transform in term of the base matrices:

$$V_3 = \prod_{r=1}^{n}\prod_{s=1}^{l-1}\exp\{-iK''W_{r+1,s}P_{r,s+1}Q_{r,s}\}\cdot \exp\{iK''W_{r+1,l}P_{r,1}Q_{r,l}U''_r\}; \quad (11)$$

$$V_2 = \prod_{s=1}^{l} \prod_{r=1}^{n-1} \exp\{-iK'P_{r+1,s}Q_{r,s}\} \cdot \exp\{iK'P_{1,s}Q_{n,s}U'_s\} ; \tag{12}$$

$$V_1 = \prod_{s=1}^{l} \prod_{r=1}^{n} \exp\{iK*\cdot P_{r,s}Q_{r,s}\} . \tag{13}$$

where U' and U'' are boundary factors, which are similar to those in the 2D Ising model[13] and the internal factors $W_{r+1,s}$ and $W_{r+1,l}$ are new and are defined as:[14]

$$W_{r+1,s} = i^{n-1}\Gamma_{r+1,s}\Gamma_{r+2,s}...\Gamma_{r-3,s+1}\Gamma_{r-2,s+1} , \tag{14a}$$

$$W_{r+1,l} = i^{n-1}\Gamma_{r+1,l}\Gamma_{r+2,l}...\Gamma_{n,l}\Gamma_{1,1}...\Gamma_{r-2,1} . \tag{14b}$$

It should be noticed that due to the symmetry of the system, the internal factors $W_{r+1,,s}$ and $W_{r+1,,l}$ can appear in either $V_2$ or $V_3$.[6,7,14] In other words, the transfer matrices $V_2$ and $V_3$ can interexchange their roles in the $V$.

Kaufman[13] made the remarkable observation that one can decompose the transfer matrices V of the 2D Ising model as the product of factors like $e^{\theta\Gamma\Gamma/2}$, which is interpreted as a 2D rotation with rotation angle θ in the direct product space. As the transfer matrices V of the 2D Ising model are decomposed into 2 pieces of subspaces in accordance with factors $\frac{1}{2}(I \pm U)$,[13] the transfer matrices V of the 3D Ising model should be decomposed into $2^l \times (2 \times 2^{nl})$ pieces,[14] in accordance with the projection operators $\frac{1}{2}(I \pm U')$, $\frac{1}{2}(I \pm U'')$ and $\frac{1}{2}(I \pm W)$. Of course, only one piece will produce the largest eigenvalue, which dominates the partition function.

It is known that the difficulties in solving exactly the 3D Ising model are topologic, which originate from the crossover of nonplanar bonds (i.e., high – order terms in the transfer matrix[6,7]). On the other hand, in the thermodynamic limit, the largest eigenvalue as well as the partition function and its thermodynamic consequences are not affected by any boundary conditions. According to the Bogoliubov inequality, the surface to volume ratio vanishes for the infinite system[6] and therefore, the details of the effects of boundary conditions (i.e., these boundary factors in eqs. (11) and (12)) can be negligible, so matrices **V₁**, **V₂** and **V₃** in eqs. (11), (12) and (13) can be reduced to:[6]

$$V_3 = \prod_{r=1}^{n} \prod_{s=1}^{l-1} \exp\{-iK''W_{r+1,s}P_{r,s+1}Q_{r,s}\} = \prod_{j=1}^{nl} \exp\{-iK''\Gamma_{2j}\left[\prod_{k=j+1}^{j+n-1} i\Gamma_{2k-1}\Gamma_{2k}\right]\Gamma_{2j+2n-1}\}; \quad (15)$$

$$V_2 = \prod_{s=1}^{l} \prod_{r=1}^{n-1} \exp\{-iK'P_{r+1,s}Q_{r,s}\} = \prod_{j=1}^{nl} \exp\{-iK'\Gamma_{2j}\Gamma_{2j+1}\}; \quad (16)$$

$$V_1 = \prod_{s=1}^{l} \prod_{r=1}^{n} \exp\{iK^* \cdot P_{r,s}Q_{r,s}\} = \prod_{j=1}^{nl} \exp\{iK^* \cdot \Gamma_{2j-1}\Gamma_{2j}\}. \quad (17)$$

Here j runs from 1 to nl, corresponding to (r, s) running from (1, 1) to (n, l); or one has j = (n - 1) r + s. These matrices with open boundary conditions greatly simplify the procedure for solving exactly the solution of the 3D Ising model. The difficulties, i.e., the internal factors $W_{r+1,s}$, remain in eq. (15). [6,7,14]

According to the conjecture 1, an additional rotation in the 2n·l·o-space appears in **V** as an additional matrix **V'₄**:[2]

$$V'_4 = \prod_{t=1}^{n \cdot l \cdot o - 1} \exp\{-iK'''P_{t+1}Q_t\}. \quad (18)$$

with:

$$K''' = \frac{K'K''}{K} \quad \text{for } K \neq 0, \tag{19}$$

Then matrices **V₁**, **V₂** and **V₃** become:[2]

$$V'_3 = \prod_{t=1}^{n \cdot l \cdot o - 1} \exp\{-iK'' P_{t+1} Q_t\}; \tag{20}$$

$$V'_2 = \prod_{t=1}^{n \cdot l \cdot o - 1} \exp\{-iK' P_{t+1} Q_t\}; \tag{21}$$

$$V'_1 = \prod_{t=1}^{n \cdot l \cdot o} \exp\{iK^* \cdot P_t Q_t\}. \tag{22}$$

In eqs. (18), (20) and (21), we have already omitted the boundary factors, using the open boundary conditions. The introduction of the new transfer matrix **V'₄** corresponds to an additional dimension, which can be realized by performing the time average. Meanwhile, three other matrices **V'₁**, **V'₂** and **V'₃** also naturally have the forms with the additional dimension. The additional rotation in the $2n \cdot l \cdot o$-space has a counterpart transformation of the spin representation in the $2^{n \cdot l \cdot o}$-space. After introducing the additional dimension, we construct $2^{n \cdot l \cdot o}$-dimensional quaternion matrices as:

$$C_t = iP_t Q_t = 1 \otimes 1 \otimes \cdots \otimes C \otimes 1 \otimes \cdots, \tag{23a}$$

$$s'_t = C_1 C_2 \cdots C_{t-1} P_t = 1 \otimes 1 \otimes \cdots \otimes s \otimes 1 \otimes \cdots, \tag{23b}$$

$$s''_t = C_1 C_2 \cdots C_{t-1} Q_t = 1 \otimes 1 \otimes \cdots \otimes isC \otimes 1 \otimes \cdots, \tag{23c}$$

The 2n·l·o-normalized eigenvectors of the 3D Ising model behave complex quaternion eigenvectors with weight factors $w_x$, $w_y$ and $w_z$ (see Eqs. (33a) and (33b) of ref. [2]). Then the transfer matrices **V'** consist of four matrices **V'$_1$, V'$_2$, V'$_3$** and **V'$_4$** can be diagonalized by the procedure with the quaternionic framework (see Eqs. (21)-(32) and Eqs. (34)-(48) of ref. [2]), however, also following the works of Onsager[1] and Kaufman[13].

The partition function, and thermodynamic consequences such as the specific heat, the spontaneous magnetization and the true range $\kappa_x$ of the correlation, and the correlation functions of the 3D Ising model are obtained,[2] based on the two conjectures. The partition function of the 3D simple orthorhombic Ising model, being dealt within a (3 + 1) - dimensional framework with weight factors on the eigenvectors, can be written as: [2,4]

$$N^{-1} \ln Z = \ln 2 + \frac{1}{2(2\pi)^4} \int_{-\pi}^{\pi}\int_{-\pi}^{\pi}\int_{-\pi}^{\pi}\int_{-\pi}^{\pi} \ln[\cosh 2K \cosh 2(K'+K''+K''') - \sinh 2K \cos \omega'$$
$$- \sinh 2(K'+K''+K''')(|w_x|\cos\phi_x \cos\omega_x + |w_y|\cos\phi_y \cos\omega_y + |w_z|\cos\phi_z \cos\omega_z)]$$
$$d\omega' d\omega_x d\omega_y d\omega_z$$

(24)

The weight factors $w_x$, $w_y$ and $w_z$ in the eigenvectors in eqn (33) of ref. [2] have been generalized as complex numbers $|w_x| e^{i\phi_x}$, $|w_y| e^{i\phi_y}$, and $|w_z| e^{i\phi_z}$ with phases $\phi_x$, $\phi_y$, and $\phi_z$.[4] Thus, the weight factors $w_x$, $w_y$ and $w_z$ in eqn (49) of ref. [2] have been replaced by $|w_x|\cos\phi_x$, $|w_y|\cos\phi_y$ and $|w_z|\cos\phi_z$, respectively,[4] since only the real part of the phase factors appears in the eigenvalues.

The spontaneous magnetization I for the 3D simple orthorhombic Ising lattices is obtained as:[2]

$$I = \left[1 - \frac{16 x_1^2 x_2^2 x_3^2 x_4^2}{(1-x_1^2)^2 (1-x_2^2 x_3^2 x_4^2)^2}\right]^{\frac{3}{8}} \quad (25)$$

where $x_i = e^{-2K_i}$ (i = 1, 2, 3, 4), where $K_i = \beta J_i$ (i = 1,2,3). The critical temperature of the simple orthorhombic Ising lattices is determined by the relation of $KK^* = KK'+KK''+K'K''$.[2] The golden ratio $x_c = e^{-2K_c} = \frac{\sqrt{5}-1}{2}$ (or silver ratio $x_c = \sqrt{2}-1$) is the largest solution for the critical temperature of the 3D (or 2D) Ising systems, which corresponds to the most symmetric lattice in 3D (or 2D). The critical exponents for the 3D Ising model were determined to be α = 0, β = 3/8, γ = 5/4, δ = 13/3, η = 1/8 and ν = 2/3, satisfying the scaling laws.

### III. OUTLOOK ON MATHEMATICAL STRUCTUIRE

In this section, an overview is given for the mathematical structures of the 3D Ising model, including algebraic aspects (Lie algebra, Clifford or geometric algebra, Jordan algebra, conformal algebra, etc.), topologic aspects, and geometric aspects. Much attention is focused on hypercomplex and Jordan-von Neumann-Wigner procedures, Yang-Baxter relations and tetrahedron relations, etc. .

#### A. Algebraic aspects and Jordan-von Neumann-Wigner procedures

The procedure for solving the 3D Ising model is related with Lie algebras/Lie group, via quaternions, Pauli matrices, special unitary group SU(2), rotation matrices $SO_2(\mathbf{R})$, and special orthogonal group SO(3). One can deal with the 3D Ising model in much larger Hilbert space by introducing the additional dimension, because the operators generate a much larger Lie algebras, due to the appearance of nonlocal behaviour (knots).[2] In 3D Ising model, one obtains a paravector by adding the fourth dimension to form quaternion eigenvectors, giving a result which corresponds to the Clifford or geometric algebra $C\ell_3$.

The algebraic approach to quantum mechanics can be based on the Jordan algebras. It is clear that the quaternion approach developed in ref. [2] can be made more elegant and simple by the use of Clifford structures and the P. Jordan structures.[9-12] The natural appearance of the multiplication $A \circ B = \frac{1}{2}(AB + BA)$ in Jordan algebras instead of the usual matrix multiplication AB satisfies the desire for commutative subalgebras of the algebra constructed in ref. [2] and for the combinatorial properties. As illustrated in Figure 2 of ref. [9], six generations of Jordan algebras are systematized by the Jordan-von Neumann-Wigner theorem.[15] It is well known that the unit quaternions can be thought of as a choice of a group structure on the 3-sphere $S^3$ that gives the group Spin(3), which is isomorphic to SU(2) and also to the universal cover of SO(3). Therefore, the quaternion basis constructed in ref. [2] for the 3D Ising model represents naturally arotation in the 4D space (a (3 + 1) – dimensional space-time), which certify the validity of the Conjecture 1.[2] The quaternionic bases found in ref. [2] for the 3D Ising model are complexified quaternionic bases, on the 3-sphere $S^3$. Performing the 4-fold integrand of the partition function of the 3D Ising model meets the requirement of taking the time average. This procedure is related closely with well-developed theories, for example, complexified quaternion,[16] quaternionic quantum mechanics,[17-19] and quaternion and special relativity.[20]

In a recent work,[21] Zhang and March proposed that quaternion-based functions developed in ref. [2] for the 3D Ising model can be utilized to study the conformal invariance in dimensions higher than two. The 2D conformal field theory can be generalized to be appropriate for three dimensions, within the framework of the quaternionic coordinates with complex weights. The 3D conformal transformations can be decomposed into three 2D conformal transformations, where the Virasoro algebra still works, but only for each complex plane of quaternionic coordinates in the complexified quaternionic Hilbert space. The local conformal invariance in 3D is limited in each complex plane of quaternionic coordinates. Then one needs to perform the summation i of the 2D conformal blocks in three complex planes together with the contributions of the phase factors $w_i$. Please notice that for this procedure, three independent Virasoro algebras with weight factors ($\mathrm{Re}\left|e^{i\phi_i}\right|$ could be zero or non-zero values) guarantee that three independent Virasoro algebras can be written within the 3+1 dimensional space, and one does not need to introduce a 6-dimensional (2+2+2) space for it. Although the origin of the complex weights (and how to derive them quantitatively) has not been understood well yet,[21] the quaternionic coordinates with complex weights provide a reliable proposal for dealing with the 3D conformal field theory.

## B. Topologic basis for a unitary transformation

The main difficulties caused by high-order terms, so-called internal factors in the transfer matrix, are topologic.[6,7] The essential ingredient in the NP-completeness of the 3D Ising model is nonplanarity,[22] indicating that the root of difficulties for solving the problem exactly is topologic. However, such NP-completeness only prevents algorithms from solving all instances of the problem in polynomial time,[22] which is insufficient to judge whether the exact solution exists. In what follows, we shall introduce briefly the knots and their relation with statistical physics to show that the conjectures introducing the fourth dimension are meaningful for dealing with the topologic problem in the 3D Ising model.

It is known that there are close connections between statistical physics and the Jones polynomial and its generalization.[23-27] The Jones polynomial of a closed braid is the partition function for a statistical model on the braid. The basic topological deformations of a plane curve are move zero and Reidemeister moves (I, II, III) for knot and link diagrams.[23-25] The Reidemeister moves change the graphical structure of a diagram while leaving the topological type of the embedding of the corresponding knot or link the same, i.e., so-called ambient isotopy. A state of the knot diagram is in analogy to the energetic states of a physical system. There is a way to preserve the state structure as the system is deformed topologically, making the invariant properties of states become topological invariants of the knot or link. The topological evolution of states and the integration over the space of states for a given system are complementary in studying the topology of knots and links. Topologically, there are two choices for smoothing a given crossing (×), and thus there are $2^N$ states of a diagram with $N$ crossings.[4,26] The bracket polynomial, i.e., the state summation, is defined by the formula:[4,23-25]

$$\langle K \rangle = \sum_{\sigma} \langle K | \sigma \rangle d^{\|\sigma\|} \tag{26}$$

Here σ run over all the states of K. d = $-A^2 - B^2$, with A, B and d being commuting algebraic variables. Note that the bracket state summation is an analog of a partition function in discrete statistical mechanics, which can be used to express the partition function for the Potts model for appropriate choices of commuting algebraic variables.[23-26]

According to the topological theory, one has:[23-25]

$$\langle \chi \rangle = A \left\langle \begin{matrix} \cup \\ \cap \end{matrix} \right\rangle + B \langle )( \rangle \tag{27a}$$

$$\langle \chi^{-1} \rangle = B \left\langle \begin{matrix} \cup \\ \cap \end{matrix} \right\rangle + A \langle )( \rangle \tag{27b}$$

The formulae (27) can be rewritten in form of a matrix as:

$$\begin{bmatrix} \langle \chi \rangle \\ \langle \chi^{-1} \rangle \end{bmatrix} = \begin{bmatrix} A & B \\ B & A \end{bmatrix} \begin{bmatrix} \left\langle \begin{matrix} \cup \\ \cap \end{matrix} \right\rangle \\ \langle )( \rangle \end{bmatrix} \tag{28a}$$

with its reverse

$$\begin{bmatrix} \left\langle \begin{matrix} \cup \\ \cap \end{matrix} \right\rangle \\ \langle )( \rangle \end{bmatrix} = \frac{1}{A^2 - B^2} \begin{bmatrix} A & -B \\ -B & A \end{bmatrix} \begin{bmatrix} \langle \chi \rangle \\ \langle \chi^{-1} \rangle \end{bmatrix}. \tag{28b}$$

It is clear that one could transform from the basis of <χ> and <χ$^{-1}$> to the basis of $\langle \begin{matrix} \cup \\ \cap \end{matrix} \rangle$ and <)(> by a transformation, and vice versa. The bracket with B = $A^{-1}$, d = $-A^2$

– $A^{-2}$ is invariant under the Reidemeister moves II and III. As long as knots or links exist in a system, a (complex) matrix representing the unitary transformation may always exist, no matter how complicated the knots or links are. This indicates that actually, one does not need to "introduce" an additional rotation as proposed in Conjecture 1, since the matrix that is the representation of such a rotation exists intrinsically and spontaneously in the system. The 3D interacting system with non-trivial knots or links requires naturally the existence of the additional dimension (say, 'time'). The analysis above certifies definitely the validity of the Conjecture 1.

Meanwhile, a diagram of a knot or link can usually be interpreted as an abstract tensor diagram, by using an oriented diagram and associating two matrices $R^{ab}_{cd}$ and $\overline{R}^{ab}_{cd}$ to two types of crossing. Then, any oriented link diagram K can be mapped to a specific contracted abstract tensor T(K). If the matrices $R$ and $\overline{R}$ satisfy channel unitary, cross-channel unitary and Yang-Baxter equation, then T(K) is a regular isotopy invariant for oriented diagram K. It is worthwhile noting that the Yang-Baxter equation corresponds to a Reidemeister Move of type III. The $R^{ab}_{cd}$ can be taken to represent the scattering amplitude for a particle interaction with incoming spins (or charges) a and b and outgoing spins (or charges) c and d. One has:[23]

$$R^{ab}_{cd} = A\delta^a_c \delta^b_d + A^{-1}\delta^{ab}\delta_{cd} \tag{29a}$$

$$\overline{R}^{ab}_{cd} = A^{-1}\delta^a_c \delta^b_d + A\delta^{ab}\delta_{cd} \tag{29b}$$

with n = $-A^2 - A^{-2}$, which is a solution to the Yang-Baxter equation.[23] $\delta^a_b$ and $\delta^{ab}$ are Kronecker deltas. T(OK) = n T(K) and T(O) = n. One can study the representation theory of the quantum groups SL(2)$_q$ as the representation theory of SL (2). That the structure of a universal R-matrix is a solution of the Yang-Baxter equation emerges from the Lie algebra formalism. The Yang-Baxter equation in form of the universal R-matrix can be represented as: [23]

$$R_{12}R_{13}R_{23} = R_{23}R_{13}R_{12} \tag{30}$$

with $R_{12} = \sum_s e_s \otimes e^s \otimes 1$, $R_{13} = \sum_s e_s \otimes 1 \otimes e^s$ and $R_{23} = \sum_s 1 \otimes e_s \otimes e^s$. The characteristic polynomial of a given linear transformation can be expressed as the trace of an associated transformation on the exterior algebra of the vector space for A. This trace can be related to statistical mechanics model of the Alexander polynomial. Therefore, it is understood that the matrix **V** for the 3D Ising model consists of two kinds of contributions:[4] those reflecting the local arrangement of spins and the others reflecting the non-local behaviour of the knots. After smoothing, there will be no crossing in the new matrix $V' \equiv V'_4 \cdot V'_3 \cdot V'_2 \cdot V'_1$,[4] which precisely includes the topologic contribution to the partition function and becomes diagonalizable. The intrinsic non-local behaviour caused by the knots requires by itself the additional rotation matrix as well as the extra dimension to handle the procedure in the much larger Hilbert space, since in 3D the operators of interest generate a much larger Lie algebra.[2,28] This merely performs a unitary transformation on the Hamiltonian and the wavevectors of the system. Clearly, any procedure (like, low- and high-temperature expansions, Monte Carlo method, renormalization groups, etc.), that takes only the local spin configurations into account (without topological contributions), cannot be exact for the 3D Ising model.[2,4,7] This is because the global (topologic) effect exists in the 3D Ising system so that the flopping of a spin will sensitively affect the alignment of another spin located far from it (even with infinite distance). This can be clearly seen from the internal factors W in Eqs. (11) or (15) above.

### C. Yang-Baxter relations and Tetrahedron equations

The Yang-Baxter equation originates in a statistical mechanics problem which demands that an R-matrix associated with a 4-valent vertex commute with the row-to-row transfer matrix for the lattice. The Yang-Baxter equation and its generalization are very important for exactly solvable models of statistical mechanics,[29] because they ensure the commutativity of the transfer matrices and the integrability of the models.[30] This is due to the fact that the local weights in a partition function are often expressed in terms of solutions to a Yang-Baxter matrix equation that turns out to fit perfectly invariance under the third Reidemeister move. The transfer matrices satisfy a functional matrix relation, which, together with the commutation properties, determines their eigenvalues.[31]

For the model with spins on sites, in particular, the Ising model, the Yang-Baxter relation becomes the star-triangle relation.[29,32,33] The star-triangle relation was first developed in electric networks,[34] showing the equivalence between three resistors arranged as a star and as a triangle in a network, also known as wye – delta (Y - Δ) transformation.

$$R_1 \overline{R_1} = R_2 \overline{R_2} = R_3 \overline{R_3} = R_1 R_2 + R_2 R_3 + R_3 R_1 = \overline{R_1}\,\overline{R_2}\,\overline{R_3}/(\overline{R_1} + \overline{R_2} + \overline{R_3}) \tag{31}$$

Onsager was aware of the star-triangle relation and the resulting commutation relations of the transfer matrices, which enabled him to calculate exactly the eigenvalues of the 2D Ising model.[1] The star-triangle relation was used by Wannier[35] and Houtappel[36] to locate the critical point of the triangular and honeycomb lattice Ising models.

The Yang-Baxter relations, corresponding to the Reidemeister move of type III in knot theory, can be represented by the defining relations of Artin's braid group.[34] A braid can be represented by a product of the operators $R_{i,i+1}$ and their inverses, provided:[34]

$$R_{i,i+1}R_{i+1,i+2}R_{i,i+1} = R_{i+1,i+2}R_{i,i+1}R_{i+1,i+2} \tag{32a}$$

and

$$[R_{i,i+1}, R_{j,j+1}] = 0, \text{ if } |i-j| \geq 2. \tag{32b}$$

It is interesting to notice that the Yang-Baxter relations also reflect a fact that the three-body S-matrix can be factorized in terms of two-body contributions since any three-body collision can be regarded as a succession of two-body collisions and the order of the collisions does not affect the final outcome.[37] In 1960's, one dimensional quantum N-body problem with delta-function interactions,[38] and the anisotropic Heisenberg spin chain[39] were solved by the Bethe Ansatz approach, and the so-called Yang-Baxter relation was observed by Yang.[40] A generalization of the Artin's braid relations (32a) and (32b) was obtained as:[34]

$$R_{i,i+1}(p-q)R_{i+1,i+2}(p-r)R_{i,i+1}(q-r) = R_{i+1,i+2}(q-r)R_{i,i+1}(p-r)R_{i+1,i+2}(p-q) \tag{33}$$

with rapidities p, q and r. The Bethe Ansatz approach was then used to solve 2D vertex models,[30,41-44] the three-spin interaction models,[45] the hard-hexagon model,[46] the Fateev-Zamolodchikov model,[47] the Kashiwara-Miwa model,[48] the

Andrews-Baxter-Forrester model,[49] the interaction-round-a-face (IRF) models,[50] the critical Potts models[50] and the chiral Potts models. The Yang-Baxter relation takes different forms in different models.[34] The star-triangle relations guarantee completely the integrability of the model and the commuting transfer matrix can be constructed as $T(u)\,T(v) - T(v)\,T(u) = 0$.

It is understood that the Yang-Baxter equation can be utilized only to solve the 2D models, and one needs so-called tetrahedron equation, or generalized Yang-Baxter equation, to deal with the 3D models. Tetrahedron equations were introduced by Zamolodchikov as a 3D generalization of the Yang-Baxter equations, who found a special solution.[51,52] The tetrahedron equation satisfied by weight functions plays an important role, which, analogous to the Yang-Baxter equation, preserves the commutativity of layer-to-layer transfer matrices constructed from the weight functions. One can deduce global properties like the commutativity of transfer matrices, by imposing the tetrahedron equation on the local statistical weights of the model. These local symmetry relations can be applied to derive the global properties of the model, since one can associate local statistical weights at all the intersections of the lattice, in such a way that the tetrahedron and inverse relations are satisfied everywhere on the lattice, and the transformations will leave the partition function unchanged.[53]

Stroganov[54] gave a survey of results on the 3D generalization of the Yang-Baxter equation, and discussed the integrability condition (i.e., tetrahedron equation) for statistical spin models on a simple cubic lattice. The 3D statistical system can be treated as a 2D system with a composite weight. The trick is that, projecting the cubic lattice along the third direction results in a quadratic lattice with the effective Boltzmann weight. A sufficient condition for two transfer matrices V and V′ to commute is the Yang-Baxter equation for the composite weights $R_{12}$ and $R'_{14}$, $R_{12}R'_{14}R''_{24} = R''_{24}R'_{14}R_{12}$. Then the composite Yang-Baxter equation will be satisfied if there exists an auxiliary nondegenerate matrix R‴ such that $l_\alpha R''' = R''' r_\alpha$ and the traces of the products of M auxiliary matrices $l_\alpha$ and $r_\alpha$ are equal.[54] The composite Yang-Baxter equation, so-called tetrahedron equation, can be written as:[54]

$$\sum_{\substack{k_1,k_2,k_3,\\k_4,k_5,k_6}} R^{k_1,k_2,k_3}_{i_1,i_2,i_3} R'^{j_1,k_4,k_5}_{k_1,i_4,i_5} R''^{j_2,j_4,k_6}_{k_2,k_4,i_6} R'''^{j_3,j_5,j_6}_{k_3,k_5,k_6}$$
$$= \sum_{\substack{k_1,k_2,k_3,\\k_4,k_5,k_6}} R'''^{k_3,k_5,k_6}_{i_3,i_5,i_6} R''^{k_2,k_4,j_6}_{i_2,i_4,k_6} R'^{k_1,j_4,j_5}_{i_1,k_4,k_5} R^{j_1,j_2,j_3}_{k_1,k_2,k_3} \quad (34)$$

or in a simplified form:[54-57]

$$R_{123}R_{145}R_{246}R_{356} = R_{356}R_{246}R_{145}R_{123} \quad (35)$$

Different versions of the tetrahedron equation were considered together with their symmetrical properties.[54]

Bazhanov and Baxter showed that this solution can be seen as a special case of the sl(n) chiral Potts model.[58,59] The partition function of a sl(n) chiral Potts model can be written in terms of traces of a layer-to-layer transfer matrix T with elements being the products of all the V functions of cubes between two adjacent layers. The transfer matrix T depends on the Boltzmann weight function V, which can be written as T(V). Two transfer matrices T(V) and T(V′) commute, $[T(V), T(V')] = 0$, if there exist two other Boltzmann weight functions U and W such that V, V′, U, and

W satisfy the tetrahedron relation.

$$V \cdot V' \cdot U \cdot W = W \cdot U \cdot V' \cdot V \tag{36}$$

Alternatively, the two transfer matrices T(V) and T(V') will commute if there exists another weight function V'' such that the Boltzmann weights S, S', and S'', each of which is of a parallelepiped Γ formed by a line of n cubes with the periodic boundary, satisfy the Yang-Baxter equation.[58]

$$\sum_{\sigma} S(\alpha,\beta,\gamma,\sigma)S'(\sigma,\gamma,\delta,\varepsilon)S''(\alpha,\sigma,\varepsilon,\kappa) = \sum_{\sigma} S(\kappa,\sigma,\delta,\varepsilon)S'(\alpha,\beta,\sigma,\kappa)S''(\beta,\gamma,\delta,\sigma) \tag{37}$$

with $S(\alpha,\beta,\gamma,\delta) = \prod_{cube \in \Gamma} V(a|e,f,g|b,c,d|h)$ (and S' and S'' with V replaced by V' and V''), a,..,h are the eight spins of the cube and $V(a|e,f,g|b,c,d|h)$ is the Boltzmann weight of the spin configuration a,..,h of the interaction-round-a-cube model. Namely, one is required to satisfy the Yang-Baxter equation for composite "two-dimensional" weights.

Decomposition and rearrangement of a tetrahedron (of a rhombic dodecahedron) show how to deal with the topological problem[53] by disconnection/fusion of the crossings. Satisfying the tetrahedron relation guarantees the commutativity of the transfer matrices and the integrability of the 3D Ising models. As shown in eq. (2) or (10), the partition function of the 3D Ising model can be written in terms of the traces of a layer-to-layer transfer matrix T with elements being the products of all the V functions of cubes between two adjacent layers. Similar to the previous results,[53,54,58,59] the two transfer matrices T(V) and T(V') will commute, i.e., $[T(V),T(V')] = 0$, if one figures out the solution of the tetrahedron relation (or the composite Yang-Baxter equation) for the 3D Ising model. However, a difficulty is that usually, such a system is overdetermined and constraints must be imposed on the variables of each local weight to allow for a solution.[53] Nevertheless, such the tetrahedron relation should exist, since the Jordan algebra and the Jordan-von Neumann-Wigner procedures have already guarantee the commutative relations.[9-12] Since the Yang-Baxter equation does not involve the disconnection/fusion of the crossings, no geometrical phase factors generalize during the procedure for the 2D Ising model without crossings. However, a tetrahedron relation does involve the disconnection/fusion of the crossings, which causes the emerging of the phase factors in the 3D Ising model.

### D. Geometric aspects

For the 2D Ising model, the geometric relations obtained in ref. [1,13] are those for a hyperbolic triangle, which are represented in the 2D Poincaré disk model. For the 3D Ising model, the geometric relations obtained in ref. [2], i.e., eqns. (29)-(32) of ref. [2], are those in hyperbolic 3-sphere (or 4-ball), which can be represented in the 4D Poincaré disk (ball) model. Notice that the 3-sphere has a natural Lie group structure given by quaternion multiplication. This interesting geometry of the 3-sphere is consistent with our idea of quaternion eigenvectors constructed in ref. [2] for the 3D Ising model. According to observations in ref. [2], the duality transformations of the simple orthorhombic Ising models are between the edges and their corresponding faces of the two dual orthorhombic lattices. Therefore, the duality between other 3D lattices should be related also with the edges and their corresponding faces of the dual lattices. It is known that a dual polyhedron of a tetrahedron with unit edge lengths is another oppositely oriented tetrahedron with unit edge lengths. One can find the duality relation between two dual tetrahedron lattices, or alternatively, between a tetrahedron lattice and a 3D honeycomb lattice. The duality relation could map a low-temperature (high-temperature) model on the tetrahedron lattice to a high-temperature (low-temperature) one on the 3D honeycomb lattice.

The condition for the critical temperature $KK^* = KK'+KK''+K'K''$ obtained in ref. [2] is actually a star-triangle relation, i.e., the (composite) Yang-Baxter equation in the continuous limit. The weight factors for the 3D Ising model are geometric phases, similar to the Berry phase effect, the Aharonov-Bohm effect, the Josephson effect, the Quantum Hall effects, etc.[4] Moreover, the balance between the exchange energy and thermal activity of the Ising model is related with the geometric duality, fractal and quasicrystals. There is the duality between a cube and its inscribed Icosahedron via the golden ratio, which is related with quasicrystals. The critical point of the cube Ising model is located at the golden ratio,[2] which indicates that the balance between the exchange energy and thermal activity is reached at this point for this most symmetric 3D lattice. Meanwhile, the critical point of the square Ising model is located at the silver ratio. The golden ratio and the silver ratio are also related to Fibonacci number and Octonacci number, respectively, which are the projecting angles of 3D 10-fold and 2D 8-fold quasicrystals. Therefore, there should be some geometrical connection between them. On the other hand, the golden ratio and the silver ratio can be represented as two kinds of fractals: of flower type and of branch

type, which indicates the duality between the two types of fractals.[2,9-12]

## IV. DISCUSSION

### A. Relativistic quantum statistical mechanics and its relation with quantum field and gauge theories

From the definition, a knot is an embedding of a circle S in a three manifold Y that is physical 3D space and S might be a superstring or cosmic string in Y. As one includes time dependence in the system, the knot S is replaced by its world sheet, a Riemann surface $\Sigma$.[60] Y is replaced by space-time, a four manifold M. Thus, to work relativistically, one will study not a knot in a three manifold system, but instead, a Riemann surface $\Sigma$ imbedded in a smooth four manifold Z. The knot theory refers to the case that $\Sigma$ is $S \times R^1$ and Z is $Y \times R^1$, with $R^1$ representing time and S a knot in the three manifold Y. In addition, according the knot theory, a knot in three dimensions can be untied when placed in a four-dimensional space, which is done by changing crossings. In fact, in four dimensions, any non-intersecting closed loop of one-dimensional string is equivalent to an unknot. Thus we could formulate a relativistic quantum field theory or a relativistic quantum statistical mechanics model, which is suitable for studying an embedding $\phi: \Sigma \to Z$. Dealing with the 3D Ising model in the (3+1) dimensional framework realizes this purpose. As mentioned above, the complexified quaternionic bases constructed above (also in ref. [2]) set up a (3+1) dimensional framework using the time average, by performing the 4-fold integrand of the partition function of the 3D Ising model. As pointed out in ref. [7], the ergodic hypothesis has been proved to be one of the most difficult problems and its proof under fairly general conditions is lacking. The lack of ergodicity of the 3D Ising model leads to the time average being different from the ensemble average.[7] With the help of the additional dimension and novel topologic phases, we actually deal with a relativistic quantum statistical mechanics model with the complexified quaternionic bases. Equivalence exists between the sum of the t slices of the 3D Ising model and the (3+1) dimensional Ising model represented in the space-time framework. The internal factor in the 3D Ising Hamiltonian of each t slice can be rearranged in the space-time framework. Moreover, the unitary transformation (as well as the rotation) for smoothing the crossings actually corresponds to the Lorentz transformation in special relativity

The Jones polynomial is closely connected not only with statistical mechanics, but also with quantum field theory. Witten uncovered the following formula:[60-62]

$$V_L(e^{2\pi i/(k+2)}) = \int_A \exp\left\{\frac{i}{\hbar}\int_{S^3} tr\left(A \wedge dA + \frac{2}{3} A \wedge A \wedge A\right)\right\} \prod_j tr\left(P \exp \oint_j A\right)[DA]. \qquad (38)$$

where A ranges over all functions from $S^3$ to the Lie algebra su(2), modulo the action of the gauge group SU(2).[27] The formulism of the Witten integral implicates invariants of knots and links, corresponding to each classical Lie algebra.[25] The Wilson loop can be introduced to denote the dependence on the loop K and the field A.

$$W_K(A) = tr\left(P \exp \oint_K A\right).$$

In 2D conformal field theory, canonical quantization on a circle S gives a physical Hilbert space $H_s$. A vector $\psi \in H_s$ is a suitable functional of appropriate fields on S, which corresponds to a local field operator $O_\psi$. In conformal field theory, there is a relation between the vector in the Hilbert space and the local operator. The 3D analog of such a relation between states and local operators can be found also, as shown in ref. [61]. However, according to the quaternion approach developed in ref. [2] for the 3D Ising model, some new features are uncovered for relation between states and local operators in the 3D systems. The physical Hilbert space obtained by quantization in 2+1 dimensions can be interpreted as the spaces of conformal blocks in 1+1 dimensions.[21,62] Analogously, the physical Hilbert space obtained by quantization in 3+1 dimensions can be interpreted as the spaces of conformal blocks in three 1+1 dimensional complex planes of the quaternion coordinates.[21] Furthermore, Atiyah[63] conjectured that the Jones knot polynomials should have a natural description in terms of Floer and Donaldson theory. This proposal was accompanied by a whole list of analogies between Floer theory and the Jones knot polynomials. The connection between Floer theory of three manifolds[64] and Donaldson theory of four manifolds[65,66] may shed lights on studying the 3D Ising model within the (3+1) dimensional framework.[67] In this way, the 3D Ising model could also be regarded as a relativistic quantum statistical mechanics model.

On one hand, the 3D spin Ising model can be translated into the 3D $Z_2$ gauge model by the Kramers-Wannier duality transformation.[68,69] The duality between the 3D spin Ising model and the 3D $Z_2$ gauge model ensures that we can study the 3D $Z_2$ gauge model using directly the results obtained for the 3D Ising model. Meanwhile, the study of the 3D Ising model may shed light on understanding various gauge theories. On other hand, the partition function of a statistical mechanics model, such as an Ising model, can be closely related to the knot polynomials by matrix elements of the braiding matrices of an associated rational conformal field theory, or alternatively, the matrix elements of the R-matrix of a quantum group.[61,70] The 3D description of the knot polynomials were given from the view point of 3D Chern-Simons gauge theory. The evaluation of the partition function of the classical lattice models can be reduced to evaluate the Wilson line expectation values in the Chern-Simons gauge theory. More specifically, the vacuum expectation values of the Wilson loops can be computed from statistical mechanics if one assigns local weights to vertex configurations in a proper way,[70] which can be represented in a formula appropriate for charges in the spin-1/2 representation of SU(2). The Jones polynomials correspond to gauge group SU(2). Actually, the conception of the Jones polynomials are closely linked to the Temperley-Lieb algebras of the statistical mechanics, which is important for understanding integrable lattice models related to the Visasoro discrete series.[61] The study of the 3D Ising model may be helpful for understanding the connection between the conformal field theory and Chern-Simons gauge theory.

Different bases can be obtained for the physical Hilbert space by different constructions. Among them, there is a base without any knots/crossings, which must be linear combinations of other bases with knots/crossings. The coefficients of linear combinations can be noted as braiding matrix. An example for such braiding matrix was given in (3.3) - (3.5) of ref. [61]. Therefore, removing knots/crossings, by Reidemeister moves in the knot theory, corresponds to a braiding matrix. The expectation of the braiding matrix, with the tetrahedron (3.7) or (4.2) of ref. [61] is proportional to a exponential factor exp(-H), here H is the lattice Hamiltonian of the statistical mechanics model. For the 3D Ising model, it is understood that the knots/crossings hidden as the internal factors in the matrixes $V_2$ and $V_3$, contribute to the expectation of a braiding matrix $V'_4$, as proposed in ref. [2]. This matrix represents the topologic effects of all the knots/crossings in the 3D Ising model. The action of this matrix is merely a unitary transformation on the bases of the Hilbert space as discussed already above.

### B. Physical significance of weight factors

The weight factors for the 3D Ising model were interpreted as geometric (or topologic) phases, [4] similar to those in Berry phase effect, Aharonov-Bohm effect, Josephson effect, etc. The novel phases appeared in the 3D Ising model[2,4] can be understood further as follows:

The 3D spin Ising model can be mapped into the 3D $Z_2$ gauge model by the Kramers-Wannier duality transformation.[68,69] In the physics of gauge theories, Wilson lines correspond essentially to the space-time trajectory of a charged particle, i.e., so-called world histories of mesons or baryons.[61] One can use fractional statistics for a particle in 2+1 dimensions, meaning that the quantum wave function changes by a phase $e^{2\pi i \delta}$ under a $2\pi$ rotation. The Chern-Simons theory was connected to theory of knots and links, which can be regularized to give invariants of three manifolds and knots.[62] The particles represented by Wilson lines in the Chern-Simons theory have fractional statistics with $\delta = n_a^2/2k$ in the Abelian theory or δ = h in the non-Abelian theory.[62] where $n_a$ is an integer (corresponding to representations of the gauge group U(1)), k is the parameter appearing in the Chern-Simons three form Lagrangian (the quantization condition require that k should be an integer for SU(N)) and h is the conformal weight of a certain primary field in 1+1 dimensional current algebra. Under a change of framing, the expectation values of Wilson lines are multiplied by a phase $\exp(2\pi i h_a)$, where $h_a$ is the conformal weight of the field. A twist of a Wilson line is equivalent to a phase, while a braiding of two Wilson lines from a trivalent vertex is also equivalent to a phase. The skein relation for Wilson lines in the defining N dimensional representation of SU(N) can be found.[61] The relation between two different 'flattened' tetrahedra with over-crossing and under-crossing of the internal lines were derived in ref. [61]. A phase related to the conformal weights of the field of the external lines appears in this relation. Because the procedure in the 3D Ising model to remove the knots/crossings involve the tetrahedron relation (or the composite Yang-Baxter equation), it is natural that a phase appears in the functions of the bases of the Hilbert space, corresponding to the disconnection/fusion of the crossings. Actually, it is a common knowledge nowadays that the transformation of a system between different (space-time) frames can bring the gauge potential (or phase factors). The unitary (Lorentz) transformation for smoothing the crossings of the 3D Ising model naturally brings the phase factors. In this sense, the two conjectures introduced in ref. [2] are self-consistent.

The phases in the 3D Ising model are also analogous to those of non-Abelian anyons in fractional quantum Hall systems, which originate from interchanging many-body interacting particles (spins).[71] In 2D, a particle loop that encircles another cannot be deformed to a point without cutting the other particles, so the notion of a winding of one particle around another is well defined. It is known that a 2D system does not necessarily come back to the same state after a nontrivial winding involved in the trajectory of two particles which are interchanged twice in a clockwise manner, because it can result in a nontrivial phase $e^{2i\phi}$. Anyons are particles with the statistical angle $\phi$ unequal to 0 and $\pi$.[71] The braiding quasiparticles cause nontrivial rotations within degenerate many-quasiparticle Hilbert space, which requires a non-Abelian braid statistic. It is known that the topological properties of 3D systems are quite different from those of 2D systems. In 3D, a process in which one particle is wrapped all the way around another is topologically equivalent to a process in which none of the particles move at all. Thus, usually, the wave function should be left unchanged by two such interchanges of particles, and only two possibilities, corresponding to bosons/fermions with symmetric/antisymmetric wave functions, exist for the case that the wave function is changed by a ± sign under a single interchange.[71] But, if a system of many particles (no matter whether they are fermions or bosons) confined to a 2D plane has excitations which are localized disturbances of its quantum mechanical ground state, then these quasiparticles can be anyons. A system is in a topological phase of matter, when it has anyonic quasiparticle excitations above its ground state.[71] The 2n·l·o-normalized eigenvectors for the 3D Ising model is analogous to 'quaternion' ones[2] with a scalar part and a 3D vector part, while those of the 2D Ising model are in the form of a scalar part and a 1D vector. The interacting particles in 3D always force themselves to be confined in one of many 2D planes and then move to another, so that the partition function of the 3D Ising model has the feature of the 2D Ising model, but with the topologic phases. This is intrinsic requirement of topological structures of the 3D Ising system. This is consistent with our discussion above for the 3D conformal transformations that can be decomposed into three 2D conformal transformations for each complex plane of quaternionic coordinates in the complexified quaternionic Hilbert space,[21] where the Virasoro algebra for 2D conformal field still works. The role of interactions in the 3D Ising model is similar to that of local "trap" potential in fractional quantum Hall systems.[71]

### C. Comparison of the exact solution with numerical study

It has been commonly acknowledged that numerical studies cannot serve as a standard for judging a putative exact solution. Note that the conjectured exact solution is obtained in thermodynamic limit ($N \to \infty$), which is suitable for either the periodic or the open boundary condition (as indicated in section II). The global (topologic) effect caused by the internal factors in transfer matrices in the 3D Ising system results in the sensitive flopping of a spin affected by the alignment of another spin located far from it (even at an infinite distance). Therefore, because of the existence of the global (topologic) effect, the results obtained by any procedure (like, low- and high-temperature expansions, Monte Carlo method, renormalization groups, etc.) that takes into account only the local spin environment and/or finite-size effect, cannot be treated as exact for the 3D Ising model. However, it is still worthwhile making a brief comparison of the conjectured exact solution in ref. [2] with numerical results (though a detailed comparison was performed in ref. [2] already).

First, let us compare the conjectured exact solution for the critical point of the 3D Ising model with the results obtained by other approximation approaches. The exact solution $v_c \equiv \tanh K_c = \sqrt{5} - 2$ for the critical point of the simple cubic Ising model coincides with the first factor of the Rosengren's conjecture obtained by analysis of a relevant class of weighted lattice walks with no backsteps,[72] while the second factor of the Rosengren's conjecture certainly has to be omitted,[73] because it was mislead by numerical calculations. Usually, the more accuracy an approach is, the lower the critical point is obtained. The putative exact solution $K_c$ = 0.24060591…… (i.e., $1/K_c$ = 4.15617384……) of the 3D simple cubic Ising model is very close to the low limit of Kikuchi's estimation, within the error of 1.6%.[74] It is located at the lowest boundary of the Oguchi' estimations within the error of ~ 0.25 %.[75-77] As it should be, the putative exact solution is lower than other approximation values obtained by various series expansion methods, such as Wakefield's method,[78,79] Bethe's first and second approximations,[79,80] Kirkwood's method,[79,81] etc.. It is slightly lower than the value of $1/K_c$ = 4.511505 in the Binder and Luijten's review,[82] which was established from the results of high-temperature series extrapolation, Monte Carlo renormalization group, Monte Carlo and finite – size scaling. It is understood here that the difference between the exact solution in ref. [2] and the numerical value in ref. [82] is attributed to the topologic terms.

The exact critical exponents based on the two conjectures for the 3D Ising lattice[2] satisfy the scaling laws and show universality behaviors. In Table 1, these exact critical exponents are compared with the approximate values obtained by the Monte Carlo renormalization group,[83] the renormalization group with the ε expansion to order $\varepsilon^2$, the high - temperature series expansion.[84-86] Nowadays, the Pelissetto and Vicari's values[83] are well – accepted by numerical calculation community, in consideration of the high – precision of simulations (but note that the high – precision does not directly imply a high accuracy since systematic errors might exist).

According to the conjectured exact solution, the specific heat of the 3D Ising lattice has the same singularity of logarithm as that of the 2D one. The small values, but non-zero, of the critical exponent α, range from 0.0625 to 0.125, obtained by the renormalization group and the high-temperature series expansion, which are attributed to the preset fitting of power laws with systematic errors (without accounting for global contributions) in those methods. Actually, the approximation approaches cannot figure out the difference between the behavior of a curve with a power law of α < 0.2 and that of logarithms.[87] The putative exact critical exponent γ of 5/4 = 1.25 for the 3D Ising model is very close to the approximation values ranging from 1.244 to 1.25. The series for the initial susceptibility at high temperatures have so far provided the smoothest and most regular patterns of behavior of coefficients, which have all been found to be positive in sign and been used to estimate the Curie temperatures and critical exponents.[86] It has been conjectured that the exact value for the critical exponent γ of the 3D Ising lattice is simply γ = 5/4.[85,86,88,89] We are quite confident that the exact critical exponent γ equals to 5/4 for the 3D Ising model. The numerical calculations for the critical exponent α (and also others with exception of γ) depend sensitively on the location of the critical point, and the critical point located by these approximation techniques is far from the exact one, thus the numerical results determined can deviate from the exact ones. Actually, in accordance with the scaling laws, all the differences between the exact critical exponents and the numerical estimates arise mainly from the determination of the critical exponent α. The critical exponent α proves considerably harder to calculate than the others, as various theoretical and experimental techniques have been tried. The differences between the putative exact solutions and the approximations are attributable to the existence of systematical errors of the Monte Carlo and the renormalization group techniques. All of these theoretical techniques similarly neglect the high-order terms, the size effects and the knots' effects, etc., all of which are very important in the 3D Ising model. The origins of these errors and the disadvantages of those approximation/numerical techniques were discussed in detail in section VIII of ref. [2]. In consideration of the insensitive dependence of the critical exponent γ to the exact location of the critical point, we suggested in ref. [2] that the numerically obtained value of the critical exponent γ is the most reliable one of the critical exponents's values determined by such approaches. Starting from the two critical exponents of α = 0 and γ = 1.2372, one easily finds β = 0.3814, δ = 4.2438, η = 0.1442 and ν = 2/3. This means that the renormalization group theory and Monte Carlo simulations are still suitable for investigating the critical phenomena. However, it is better to focus only on highly accurate determination of the critical exponent γ, and the estimation of topologic contributions related with other critical exponents.

In Binder and Luijten's review,[82] the values of $y_t = 1/\nu = 1.588(2)$ and $y_h = 3 - \beta/\nu = 2.482(2)$ are established, in accordance with data in various references, and very close with the putative exact solutions $y_t = 1/\nu = 1.5$ and $y_h = 3 - \beta/\nu = 2.4375$, within the errors of 5.87% and 1.83% respectively. The putative exact solutions are in very good agreements with the values for critical exponents $\beta$ and $\delta$ collected in Vicentini-Missoni's review,[90] which were derived by analysis of the good data in the critical region. The exact critical exponents coincide with the critical indices $\gamma$, $\delta$, $\eta$ and $\nu$ of the real fluids,[90] and ones for the interfacial tension in the two phase fluids.[91] Recently, Zhang and March found in ref. [92] that critical exponents in some bulk magnetic materials indeed form a 3D Ising universality, in good agreement with the exact solution reported in ref. [2].

To end this subsection, we emphasize once again that the results of the approximation methods cannot serve as the only standard for judging the correctness of the putative exact solutions, but the exact solution can serve in evaluating the systematical errors (i.e., the topologic contributions in the 3D Ising case) of the approximations.

### D. Singularities at/near infinite temperature

Up to date, the debates[3-8,93,94] have focused mainly on whether the high-temperature series is the standard for judging the correctness of the exact solution. The main objections[3,5,6,8,93] to our solution are based on a misjudgment that the exact solution of the 3D Ising model must pass the series test. Such an argument is based on a belief that the theorems have been rigorously established for the convergence of the high-temperature series of the Ising model. However, as pointed out already in ref. [4,7,94], all the well-known theorems[95-106] cited in refs. [3,5,6,8,93] for the convergence of the high-temperature series of the 3D Ising model are rigorously proved only for $\beta$ (= $1/k_B T$) > 0, not for infinite temperature ($\beta$ = 0).

For instance, Lebowitz and Penrose indicated clearly in the abstract of their paper[99] that their proof for the analytics of the free energy per site and the distribution function of the Ising model is for $\beta$ > 0. They pointed out in p. 102 of ref. [99] that there is no general reason to expect a series expansion of *p* or *n* in powers of $\beta$ to converge, since $\beta$ = 0 lies at the boundary of the region E of ($\beta$, z) space. Their proof for the Ising model is related with the Yang–Lee Theorems [107,108] for $\beta$ > 0, and for the analytic of the function $\beta p$.

Here, we have to inspect the definition of the hard-core model and the Ising model to discern the difference between them, and also to clarify the conditions of the proof of Lebowitz and Penrose [99] for the hard-core model. The hard-core potential is defined by $\varphi(r) = +\infty$ for $r \leq a$, and $\varphi(r) < \infty$ for $r > a$, where a is a positive constant (a > 0) [99]. The Ising ferromagnet is isomorphic to a lattice gas with an attractive interaction potential with $\varphi(0) = +\infty$, and $\varphi(r) \leq 0$ for $r \neq 0$ [99]. So, a = 0 for the Ising lattice (see also the definition (9) in page 411 of ref. [108]). The key distinction between the two models is whether a is zero or a positive constant. Though Lebowitz and Penrose claimed that the hard-core systems are analytic in $\beta$ at $\beta$ = 0, actually, their proof concerns $\beta p$ (the series (4) of ref. [99]), not p itself (For equivalence between βp and p, setting β = 1 equalizes to T = $1/k_B$ ≠ ∞; it corresponds to β ≠ 0)). This indicates clearly that actually, the hard-core systems are analytic in β only at β > 0.

Lebowitz and Penrose at the end of the section II of ref. [99] used a word of 'implies' to refer to Gallavotti et al.'s work,[102,105] and in section IV referred again to Gallavotti et al.'s work.[105] However, although Gallavotti et al. proved that the radius of convergence is greater than zero, but once again their proof does not touch β = 0.[102,105,106] In the second page (p. 275) of their paper[102] for a detailed proof, they put, for convenience, β = 1. When they defined $Z_\Lambda(\Phi)$ in eq. (5) of ref. [106], they also set β = 1. This condition of β = 1 is contradictory with β = 0 (For an equivalent between βf and f, setting β = 1 equalizes to T = $1/k_B$ ≠ ∞). The inequality $\sum_{\substack{T \cap X - \phi \\ T \neq \phi}} |K_{\beta\phi'}(X,T)| \leq \left[\exp\left(e^{\beta\|\phi'\|} - 1\right) - 1\right]$ of ref. [105] (or Proposition 1 (i.e., eq. (17)) of ref. [102]) is invalid for β = 0. Note that the summation above would be less than or equal to zero as $\|\phi'\| < +\infty$ (the condition in ref. [105]) and β = 0; if one used $\|\phi'\| = +\infty$ and β = 0, there would be uncertainty (i.e., multiply zero by infinite) for the summation; if one used $\|\phi'\| < +\infty$ and β → 0, one would still meet the uncertainty problem.

Perk claimed in his recent article[93] that his Theorem 2.9 proves rigorously the analyticity of the reduced free energy βf in terms of β at β = 0. Actually, some mathematical tricks had been performed in his procedure to avoid the difficulty of singularities at β = 0, which first appear in Definition 1.4, defining the free energy per site $f_N$ and its infinite system limit f by eqn (6), but in form of -β$f_N$.[93] Then Lemma 2.5 went on perpetrating the fraud, discussing the singularity of β$f_N$, and finally to

prove Theorem 2.9 'rigorously' for $\beta f$.[93]

We have pointed out[94] that the singularities of the reduced free energy $\beta f$, the free energy per site f and the free energy F of the 3D Ising model differ at $\beta = 0$. This is because there is a singularity at $T = \infty$, which is inconsistent with the assumption for the definition of the free energy per site f, and therefore, such a definition loses physical significance at $T = \infty$.[94,107-110] At $\beta = 0$, one has to face directly the total free energy F to study the singularities of the system. In addition to the roots of the partition function Z, one should also discuss the roots of $Z^{-1}$. Lee and Yang[107,108] discussed only the zeros of the partition function Z, since they were interested mainly in singularities (i.e., phase transition) at finite temperature. It is understood here that one needs to discuss not only the zeros, but also the poles of the partition function Z for the complete information of the system, specially, for the singularities at infinite temperature. This is because both the singularities of $Z = 0$ and $Z = \infty$ have the same character, except for a minus sign, and considerable interest should be paid to both of them. In Definition 1.4 of ref. [93], the negative sign was moved to the left-hand-side of eqn (6), to avoid the discussion on zeros with regard to $Z^{-1}$. But, if one conceals the singularities of $\ln Z^{-1}$ by mathematical tricks, one must find similar tricks to obscure the singularities of $\ln Z$ and also to violate the Lee-Yang Theorems.[107,108] Thus, the intrinsic characters of the singularities of the zeros (and the poles) at infinite temperature are quite different from those at finite temperatures, which cannot be disregarded by the usual process of removing the singularity at finite temperatures by using $-\beta f$.

In the 3D Ising model there indeed exist three singularities:[7,94] 1) H = 0, β = $β_c$; 2) H = ±i∞, β → 0; 3) H = 0, β = 0. The third singularity has physical significance:[7,94] The 3D Ising system experiences a change from a 'non-interaction' state at β = 0 to an interacting state at β > 0. This change of state is similar to a switch turning off/on all the interactions at/near infinite temperature, resulting in a change of the topologic structures and the corresponding phase factors.[2,4,7,94] The singularities of the free energy F and the free energy per site f at β = 0 support that two different forms for infinite temperature and finite temperatures could exist for the high-temperature series expansions of the free energy per site f, as revealed in ref. [2].

## V. CONCLUSIONS AND OPENING PROBLEMS

We have represented an overview of the mathematical structure of the 3D Ising model. We have convinced ourself of the following facts:

The topological problem of the 3D Ising model can be dealt with by a unitary transformation with a matrix being a spin representation in $2^{n \cdot l \cdot o}$-space for a rotation in 2n·l·o-space, which serves to satisfy the intrinsic requirement of smoothing all the crossings in the transfer matrices. Meanwhile, this matrix represents the contribution of the non-local behaviour of the knots in the partition function of the 3D Ising model. The unitary transformation for smoothing the crossings in the transfer matrices changes the wave functions by complex phases $\phi_x$, $\phi_y$, and $\phi_z$.

In order to solve rigorously the 3D Ising model, we must actually deal with relativistic quantum statistical mechanics. The multiplication $A \circ B = \frac{1}{2}(AB + BA)$ in

Jordan algebras satisfies the desire for commutative subalgebras of the algebra constructed in ref. [2] and for the combinatorial properties of the 3D Ising model. The complexified quaternionic bases constructed for the 3D Ising model take the time average into account within the Jordan-von Neumann-Wigner procedure in relation with quaternionic quantum mechanics and special relativity. Meanwhile, a tetrahedron relation (or a composite Yang-Baxter equation) will also ensure the commutativity of the transfer matrices and the integrability of the 3D Ising models, although this problem has not been fixed yet.

We have also compared with the exact solution with numerical results. The exact solution for the critical exponent $\gamma$ is in excellent agreement with that obtained by the renormalization group theory and Monte Carlo simulations, which means that these approximation approaches are suitable for the high-accurate determination of the critical exponent $\gamma$. The deviation of the numerical calculations for the critical exponent $\alpha$ (and also others with exception of $\gamma$) from the exact ones are mainly due to the technical difficulties (and the systematic errors) of those approximation techniques. The difference between the exact solution and the numerical values can be used to evaluate the systematic errors of those approximation approaches, which neglect the topologic contributions of the 3D Ising model.

Furthermore, we have discussed the singularities at/near infinite temperature. We have pointed out that the singularities of the reduced free energy $\beta f$, the free energy per site $f$ and the free energy $F$ of the 3D Ising model differ at $\beta = 0$. The rigorous proof presented in the well-known theorems,[95-106] the previous

Comments/Rejoinders[3,5,6,8] and the Perk's recent Comment[93] on the convergence of the high-temperature series expansion are only for the analyticity of the reduced free energy βf (or βp), which loses its definition at β = 0. The analyticity in β of both the hard-core and Ising models has been proved for β > 0, not for β = 0. Therefore, all of these objections lose the mathematical basis, which are thoroughly disproved. The high-temperature series expansions cannot serve as a standard for judging the correctness of the exact solution of the 3D Ising model.

There has been no comment on the topology-based approach developed in ref. [2], however, a rigorous proof is still to be pursued. The open problems in this field are as follows: 1) To find and solve the tetrahedron relation (or the composite Yang-Baxter equation); 2) to prove the geometric relation for the tetrahedron – (3D) honeycomb duality; 3) to prove rigorously the relation $K''' = \frac{K'K''}{K}$ among the rotations K, K', K'' and K''' for the 3D simple orthorhombic Ising model. If one answers the first two problems, one would solve explicitly the tetrahedron and 3D-honeycomb Ising lattices, and one would also prove the commutativity of the transfer matrices and the integrability of the simple orthorhombic Ising lattices. A recent study on topological aspects of fermions on hyperdiamond might be helpful for understanding the 3D-honeycomb Ising lattices.[111] Then the third problem would be the only open question remaining for a rigorous proof of the simple orthorhombic Ising lattice. One might need to develop novel mathematical tools to obtain a rigorous proof of the equivalence between the transfer matrixes V and V' (before/after the unitary transformation), which are respectively with and without the internal factors, and meanwhile, one has to give a detailed analysis of the topologic factors. In a parallel direction, of course, one could work on applications in various fields with the conjectured exact solution and the approach developed in ref. [2] for the simple orthorhombic Ising lattices [for examples, as in refs. [9-12, 21, 92, 112-125]].

**Acknowledgements**

I acknowledge the support of the National Natural Science Foundation of China under grant number 50831006, and thank Prof. N.H. March for helpful discussion on the analogy of topologic phases with those of non-Abelian anyons in fractional quantum Hall systems.

Table 1 Comparison of the putative exact critical exponents for the 3D Ising lattice,[2] with the approximate values obtained by the Monte Carlo renormalization group (PV-MC),[83] the renormalization group (RG) with the ε expansion to order $\varepsilon^2$, the high - temperature series expansion (SE).[84-86] PV taken from Pelissetto and Vicari's review,[83] WK from Wilson and Kogut's review,[84] F from Fisher's series expansion.[85] Notice that Domb's values in ref. [86] are same as Fisher's (T > $T_c$).[85]

| Method | α | β | γ | δ | η | ν |
|---|---|---|---|---|---|---|
| Exact | 0 | $\frac{3}{8}$ | $\frac{5}{4}$ | $\frac{13}{3}$ | $\frac{1}{8}$ | $\frac{2}{3}$ |
| WK-RG | 0.077 | 0.340 | 1.244 | 4.46 | 0.037 | 0.626 |
| PV-MC | 0.110 | 0.3265 | 1.2372 | 4.789 | 0.0364 | 0.6301 |
| WK-SE | 0.125 | 0.312 | 1.250 | 5.150 | 0.055 | 0.642 |
| F(T>$T_c$) | $\frac{1}{8}$ | $\frac{5}{16}$ | $\frac{5}{4}$ | 5 | 0 | $\frac{5}{8}$ |